\documentclass[12pt]{iopart}

\usepackage{bm}

\usepackage{tikz}
\usetikzlibrary{arrows.meta}

\newtheorem{theorem}{Theorem}

\begin{document}

\title{Average Length of Cycles in Rectangular Lattice}
\author{Ryuhei Mori}
\address{School of Computing, Tokyo Institute of Technology, Tokyo 152-8552, Japan.}
\ead{mori@c.titech.ac.jp}

\begin{abstract}
We study the number of cycles and their average length in $L\times N$ lattice by using classical method of transfer matrix.
In this work, we derive a bivariate generating function $G_3(y, z)$ in which a coefficient of $y^i z^j$ is the number of
cycles of length $i$ in $3\times j$ lattice.
By using the bivariate generating function, we show that the average length of cycles in $3\times N$ lattice is
$\alpha N + \beta + o(1)$ where $\alpha$ and $\beta$ are some algebraic numbers approximately equal to 3.166 and 0.961, respectively.
We argue generalizations of this method for $L\ge 4$, and obtain a generating function of the number of cycles
in $L\times N$ lattice for $L$ up to 7.
\end{abstract}
\maketitle

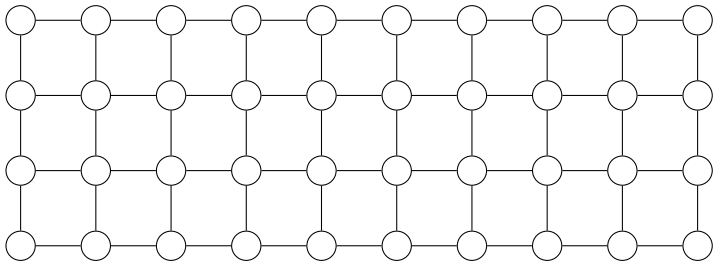
\begin{figure}[t]
\begin{center}
\begin{tikzpicture}
\foreach \x in {0,...,3}{
  \foreach \y in {0,...,9}{
    \node[circle,draw] (\y \x) at (\y, \x) {};
    \ifnum\y>0
      \pgfmathparse{int(\y-1)}
      \draw (\y \x) -- (\pgfmathresult \x);
    \fi
    \ifnum\x>0
      \pgfmathparse{int(\x-1)}
      \draw (\y \x) -- (\y \pgfmathresult);
    \fi
  }
}
\end{tikzpicture}
\end{center}
\caption{$3\times 9$ lattice.}
\label{fig:grid}
\end{figure}

\section{Introduction}
Self-avoiding walks and self-avoiding polygons on two-dimensional lattice has been studied in chemistry, physics, combinatorial mathematics and computer science~\cite{0305-4470-13-12-021, 0305-4470-38-42-001}.
In this work, we use a classical method of transfer matrix for counting the number of cycles (self-avoiding polygons) and computing their average length on $L\times N$ lattice where $L$ is fixed to be constant.
The $L\times N$ lattice has $L+1$ nodes in the vertical direction and $N+1$ nodes in the horizontal direction like $3\times 9$ lattice in \Fref{fig:grid}.
By using both of theory of generating function and numerical calculation on a computer, we obtain the following theorem.

\begin{theorem}\label{thm:main}
The average length of cycles in $3\times N$ lattice is $\alpha N+\beta+o(1)$
where $\alpha$ and $\beta$ are some algebraic numbers approximately equal to $3.166$ and $0.961$, respectively.
\end{theorem}
For obtaining Theorem~\ref{thm:main}, we introduce a bivariate generating function $G_3(y,z)$ in which a coefficient of $y^iz^j$ is the number of cycles of length $i$ in $3\times j$ lattice.
The derivation of the bivariate generating function $G_3(y,z)$ is the theoretical contribution of this work.

This paper is organized as follows.
In \Sref{sec:c3n}, we apply the classical method of transfer matrix for counting the number of cycles in $3\times N$ lattice, and obtain a generating function of this problem.
Although this is a simple extension of Stoyan and Strehl's work for Hamiltonian cycles~\cite{stoyan1996Hamiltonian}, the generating function for this problem has not been written in any literature in author's knowledge.
In \Sref{sec:l3n}, we derive the bivariate generating function $G_3(y,z)$ mentioned above.
By using this bivariate generating function, we obtain a univariate generating function of the sum of length of cycles in $3\times N$ lattice.
From the univariate generating functions, we obtain an asymptotic behavior of average length of cycles in Theorem~\ref{thm:main}.
In \Sref{sec:cln}, we argue how this method can be generalized for general $L\times N$ lattice.
The generating functions of the number of cycles in $L\times N$ lattice are shown for $L\in\{4, 5\}$ (In fact, we obtain the generating function up to $L=7$).
In \Sref{sec:sum}, we summarize this work and introduce other applications of the bivariate generating function.

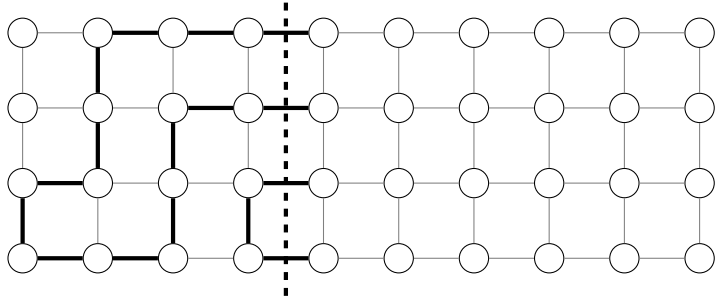
\begin{figure}[t]
\begin{center}
\begin{tikzpicture}
\foreach \x in {0,...,3}{
  \foreach \y in {0,...,9}{
    \node[circle,draw] (\y \x) at (\y, \x) {};
    \ifnum\y>0
      \pgfmathparse{int(\y-1)}
      \draw[gray] (\y \x) -- (\pgfmathresult \x);
    \fi
    \ifnum\x>0
      \pgfmathparse{int(\x-1)}
      \draw[gray] (\y \x) -- (\y \pgfmathresult);
    \fi
  }
}
\draw[ultra thick] (42) -- (32) -- (22) -- (21) -- (20) -- (10) -- (00) -- (01) -- (11) -- (12) -- (13) -- (23) -- (33) -- (43);
\draw[ultra thick] (41) -- (31) -- (30) -- (40);
\draw[ultra thick, dashed] (3.5, -0.5) -- (3.5, 3.5);
\end{tikzpicture}
\end{center}
\caption{A state on $3\times 9$ lattice. This state is represented by (1,2)(3,4).}
\label{fig:state}
\end{figure}

\section{Cycles in $3\times N$ lattice}\label{sec:c3n}
In this section, we consider the number of cycles in $3\times N$ lattice.
In the classical transfer matrix method, we focus on four horizontal edges as in \Fref{fig:state}.
A state represents chosen edges in these four edges and their connectivity by edges chosen in previous stages.
Since we are interested in cycles, the number of chosen horizontal edges in arbitrary column must be even, i.e., 0, 2 or 4.
We use a notation ($i$,$j$) for representing a state using $i$th edge and $j$th edge from top.
We also use notations (1,2)(3,4) for representing a state using all of four edges in which two top edges (and two bottom edges) are connected like \Fref{fig:state}.
The notation (1,4)(2,3) are used for representing a state using all of four edges in which top and bottom edges (and two middle edges) are connected.
In total, there are eight states except for initial state and final state. Their transition rule is shown in \Fref{fig:dp}.
Additionally, there are initial state $s$ and final state $t$ both of which represent absence of chosen edges.
The initial state $s$ connects to all of 8 states except for (1,4)(2,3) since (1,4)(2,3) cannot be generated in a single step.
The final state $t$ is connected from all of 8 states except for (1,2)(3,4) since if (1,2)(3,4) is closed, two disjoint cycles are generated.
Both of $s$ and $t$ have self-loop.
Then, the number of walks from $s$ to $t$ of length $N+1$ is the number of cycles, denoted by $C_3(N)$, in $3\times N$ lattice.
By using $10\times 10$ adjacency matrix $A$ of this directed graph, we obtain $C_3(N) = (A^{N+1})_{s,t}$.
Then, the generating function $g_3(z)$ of $C_3(N)$ can be represented by
\begin{equation*}
g_3(z) := \sum_{N=1}^\infty C_3(N) z^N
= \sum_{N=1}^\infty (A^{N+1})_{s,t} z^N
= \frac1z \left((I - z A)^{-1}\right)_{s,t}.
\end{equation*}

\begin{figure}[t]
\begin{center}
\begin{tikzpicture}[scale=.7, line width = 0.5mm, inner sep=0.4mm, node font = \small, >={Stealth[length=10pt]}]
  \node[circle,radius=50cm,draw] (n14) at ( 0, 0) {(1, 4)};
  \node[circle,draw] (n23) at ( 0, 4) {(2, 3)};
  \node[circle,draw] (n13) at (-4, 4) {(1, 3)};
  \node[circle,draw] (n24) at ( 4, 4) {(2, 4)};
  \node[circle,draw] (n12) at (-4, 0) {(1, 2)};
  \node[circle,draw] (n34) at ( 4, 0) {(3, 4)};
  \node[circle,draw] (n1234) at (-2, -4) {(1, 2)(3, 4)};
  \node[circle,draw] (n1423) at ( 2, -4) {(1, 4)(2, 3)};
  \draw[<->, very thick] (n14) to node[right=1mm] {4} (n23);
  \draw[<->, very thick] (n14) to node[above=1mm] {3} (n13);
  \draw[<->, very thick] (n14) to node[above=1mm] {3} (n24);
  \draw[<->, very thick] (n14) to node[above=1mm] {4} (n12);
  \draw[<->, very thick] (n14) to node[above=1mm] {4} (n34);
  \draw[->, very thick] (n14)  to node[above=4mm,right=.0mm] {5} (n1423);
  \draw[->, very thick] (n1234) to node[above=4mm,left=.1mm] {3} (n14);
  \draw[<->, very thick] (n23) to node[above=1mm] {3} (n13);
  \draw[<->, very thick] (n23) to node[above=1mm] {3} (n24);
  \draw[<->, very thick] (n12) to node[left=1mm] {3} (n13);
  \draw[<->, very thick] (n34) to node[right=1mm] {3} (n24);
  \draw[<-, very thick] (n1234) to node[left=1mm] {5} (n12);
  \draw[<-, very thick] (n1234) to node[very near end,left=2mm] {5} (n34);
  \draw[->, very thick] (n1423) to node[near end,below=1mm] {3} (n12);
  \draw[->, very thick] (n1423) to node[right=1mm] {3} (n34);
  \draw[<->, very thick] (n13) to [bend left=35] node[above=1mm] {4} (n24);
\end{tikzpicture}
\end{center}
\caption{Connections of the 8 states. All of 8 states have self-loop. Edge weights represent the number of edges used for the transition, which are used in~\Sref{sec:l3n}.}
\label{fig:dp}
\end{figure}

We obtain an explicit form of the generating function $g_3(z)$.
\begin{equation*}
g_3(z) = \frac{{\left(z^{5} - 2 \, z^{4} - 16 \, z^{3} + 15 \, z^{2} - 14 \, z + 6\right)} z}{{\left(2 \, z^{5} + 3 \, z^{4} - 7 \, z^{3} + 12 \, z^{2} - 7 \, z + 1\right)} {\left(z - 1\right)}^{2}}.
\end{equation*}
Here, the denominator of $g_3(z)$ is equal to
\begin{equation*}
2 \, z^{7} - z^{6} - 11 \, z^{5} + 29 \, z^{4} - 38 \, z^{3} + 27 \, z^{2} - 9 \, z + 1
\end{equation*}
which gives a linear recurrence equation
\begin{eqnarray*}
\fl
C_3(N) = 9 C_3(N-1) - 27 C_3(N-2) + 38 C_3(N-3) - 29 C_3(N-4) + 11 C_3(N-5)\\
 + C_3(N-6) - 2 C_3(N-7)
\end{eqnarray*}
for $N\ge 7$.
From the first 7 values
\begin{equation*}
[C_3(0), \cdots, C_3(6)] = [0,6,40,213,1049,5034,23984]
\end{equation*}
the sequence $\{C_3(N)\}_N$ is determined.

A generating function is also useful for analyzing an asymptotic behavior of sequences~\cite{Flajolet:2009:AC:1506267}.
Let $\lambda$ be an inverse of absolute value of the root closest to the origin of the polynomial $2z^5+3z^4-7z^3+12z^2-7z+1$, which appears at the denominator of $g_3(z)$.
Numerically, $\lambda\approx 4.75348116077213$.
Since this root is not duplicated, $C_3(N) = c \lambda^{N}(1+O(\epsilon^N))$ where $\epsilon$ is some constant smaller than 1 and
\begin{equation*}
c := \lim_{z\to1/\lambda} g_3(z) \left(1-\lambda z\right)
\approx 2.08045929462431.
\end{equation*}

\section{Length of cycles in $3\times N$ lattice}\label{sec:l3n}
For computing a sum of length of cycles, we introduce a $10\times 10$ matrix $A(y)$ which is similar to $A$ in the previous section,
but has elements $y^i$ instead of 1 where $i$ denotes the number of edges used for the transition of states.
\Fref{fig:dp} shows the number of edges used for transitions.
Of course, the number of used edges regarding self-loop, initial state and final state should be counted appropriately as well.
Then, we obtain a bivariate generating function
\begin{equation*}
G_3(y, z) = \frac1z\left((I- zA(y))^{-1}\right)_{s,t}
\end{equation*}
in which a coefficient of $y^i z^j$ is the number of cycles of length $i$ in $3\times j$ lattice.
This bivariate generating function can be computed immediately by using SageMath~\cite{sagemath}.
Explicit form of $G_3(y,z)$ is shown in Appendix.
Then, $\left.\partial G_3(y, z)/ \partial y\right|_{y=1}$ gives a univariate generating function $\ell_3(z)$ of the sum of length of cycles, denoted by $L_3(N)$, in $3\times N$ lattice.
Explicit form of this generating function is
\begin{eqnarray*}
\fl
\ell_3(z) =
-2\, z \, \Bigl(2 \, z^{11} + 2 \, z^{10} - 25 \, z^{9} + 14 \, z^{8} + 88 \, z^{7} - 300 \, z^{6} + 774 \, z^{5} - 812 \, z^{4} + 481 \, z^{3}\\
 - 250 \, z^{2} + 98 \, z - 16\Bigr) / \Bigl({\left(2 \, z^{5} + 3 \, z^{4} - 7 \, z^{3} + 12 \, z^{2} - 7 \, z + 1\right)}^{2} {\left(z - 1\right)}^{2}\Bigr)
\end{eqnarray*}
which was obtained immediately by SageMath as well.
Interestingly, the denominator of $\ell_3(z)$ is very similar to the denominator of $g_3(z)$.
Of course, the set of roots of the denominator of $\ell_3(z)$ is the same as those of $g_3(z)$.
The root closest to the origin has multiplicity 2.
In this case~\cite{Flajolet:2009:AC:1506267}, from a partial fraction decomposition of $\ell_3(z)$
\begin{eqnarray*}
\ell_3(z) &= \frac{0.965028460415963}{z-1/\lambda} + \frac{0.2915360175191517}{(z-1/\lambda)^2} + \cdots\\
&= -0.965028460415963\lambda \sum_{i=0}^\infty (\lambda z)^i\\
&\quad +  0.2915360175191517\lambda^2 \sum_{i=0}^\infty (i+1) (\lambda z)^i + \cdots
\end{eqnarray*}
given by SageMath~\cite{sagemath} and from some elementary calculations, we obtain $L_3(N) = (\alpha' N + \beta') \lambda^N(1+O(\epsilon^N))$ where $\epsilon$ is some constant smaller than 1 and
\begin{eqnarray*}
\alpha' &\approx 6.58742632385394\\
\beta' &\approx 2.00018171765772.
\end{eqnarray*}
Hence, the average length of cycles in $3\times N$ lattice is asymptotically $(\alpha' N+ \beta')/c + O(N\epsilon^N) \approx 3.16633271358645 N + 0.961413531534109 + O(N\epsilon^N)$, which
proves Theorem~\ref{thm:main}.

\section{Generalization for $L\times N$ lattice}\label{sec:cln}
\subsection{Number of states and matrix}
In this section, we consider generalizations of results in previous sections for general $L\times N$ lattice.
First, we analyze the number of states required for the transfer matrix method.
In fact, the Motzkin number plus 1 states are sufficient~\cite{stoyan1996Hamiltonian, 0305-4470-38-42-001}.
Here, the $n$th Motzkin number is the number of ways for making (not necessarily perfect) pairs on $\{1,2,\cdots,n\}$ avoiding ``alternate pairs'', i.e., $(x_1,x_2)$ and $(y_1,y_2)$ such that $x_1<y_1<x_2<y_2$.
It is easy to check that the $n$th Motzkin number $M(n)$ satisfies a recurrence equation
\begin{equation*}
M(n) = M(n-1) + \sum_{i=2}^n M(i-2) M(n-i).
\end{equation*}
Note that $M(n)=\Theta(3^n/n^{1.5})$~\cite{Flajolet:2009:AC:1506267}.

In the computation of the adjacency matrix, a decision of direct connectivity of two given states seems to be hard.
A rather simple way to construct the adjacency matrix is to apply $2^L$ choices of vertical edges for given state.
Much of them would be inconsistent, and some of them allow transition to another state.
This means that the row and column weights of the matrix is at most $2^L$.
Since the matrix is sparse, a representation of the directed graph by adjacency list is much efficient than a representation by adjacency matrix.
In this way, we could construct the adjacency lists up to $L=14$.
We obtained generating functions $g_L(z)$ for the number of cycles in $L\times N$ lattice up to $L=7$ by using SageMath~\cite{sagemath}.
Explicit forms of $g_4(z)$ and $g_5(z)$ are shown in Appendix.
We expect that we can also obtain the bivariate generating function $G_L(y, z)$ and the generating function $\ell_L(z)$ of the sum of length of cycles 
in $L\times N$ lattice for $L\ge 4$.

\subsection{Reducing the number of states}
We can reduce the number of states by using symmetricity of the problem.
For instance, there are two pairs of assymmetric states ``(1,2) and (3,4)'', and ``(1,3) and (2,4)'' for $3\times N$ lattice.
The number of configurations corresponding to (1,2) (and (1,3)) and that for (3,4) (and (2,4), respectively) are equal.
In general, for asymmetric state $u$, the number of configurations corresponding to a state $u$ and the number of configurations corresponding to a state $\bar{u}$ which is a symmetric counterpart of $u$ are equal.
It is also easy to check that the number $S(n)$ of symmetric states satisfies a recurrence equation
\begin{equation*}
S(n) = 2S(n-2) + \sum_{i=2}^{\lfloor \frac{n}2\rfloor} M(i-2) S(n-2i).
\end{equation*}
Then, we obtain an upper bound $1 + S(n) + (M(n)-S(n))/2$ of the number of required states.
\Tref{tbl:ub} shows the exact minimum number of states, which is the degree of denominator of generating function $g_L(z)$, and the upper bound obtained above.
The upper bound seems to be tight at least up to $L=7$.

The computation of generating function by SageMath from $(M(L)+1)\times(M(L)+1)$ matrix for large $L$ seems to be hard
since the size of matrix grows exponentially with respect to $L$, since each element of matrix includes a variable $z$, and since coefficients in generating function are greater than $2^{64}$.
In~\cite{stoyan1996Hamiltonian}, it is noted that the derivation of the linear recurrence relation of length $d$ on $\mathrm{GF}(p)$ from given sequence of length $2d$ by Berlekamp--Massey algorithm for several large primes $p$
and the reconstruction of the linear recurrence relation by Chinese remainder theorem gives an efficient algorithm, which takes $O(d^2)$ time for each prime.
This method may allow the computation of the linear recurrence relation up to $L=11$ (Note that $M(12)=15511$).
However, for $L=12$, a square of $M(13)=41835$ time seems to be infeasible.
We may be able to compute the generating function and the linear recurrence relation for $L\ge 12$ by exploiting the sparsity of the matrix~\cite{wiedemann1986solving}.

\begin{table}
\caption{\label{tbl:ub}Minimum number of states and upper bound.}
\begin{indented}
\item[]\begin{tabular}{@{}llll}
\br
$L$&Size of linear recurrence&Upper bound&$M(L+1)+1$\\
\mr
3&7&8&10\\
4&13&14&22\\
5&33&33&52\\
6&67&71&128\\
7&168&180&324\\
\br
\end{tabular}
\end{indented}
\end{table}

\section{Summary}\label{sec:sum}
We derive a bivariate generating functions $G_3(y, z)$ in which a coefficient of $y^i z^j$ is the number of cycles of length $i$ in $3\times j$ lattice.
We obtained a generating function of the sum of length of cycles by $\left.\partial G_3(y,z)/\partial y\right|_{y=1}$.
From analysis of generating functions, we obtain the average length $3.166 N + 0.961$ of cycles in $3\times N$ lattice.

The bivariate generating function $G_3(y,z)$ would be useful for many applications.
For instance, $\left.\partial^2 G_3(y,z)/\partial y^2\right|_{y=1}$ gives a generating function for the sum of ``length $\times$ (length $-$ 1)'' of cycles in $3\times N$ lattice.
This gives the variance of length of cycles in $3\times N$ lattice.
This method of the bivariate generating function can be generalized to larger $L$ and also to self-avoiding walks.

\ack
This work was supported by JSPS KAKENHI Grant Number JP17K17711.
The author thanks Ryo Wakatabe for correcting an error in the derivation of $A(y)$.

\section*{Appendix}

In this appendix, several generating functions are shown.
The numerator of $G_3(y,z)$ is
\begin{eqnarray*}
&-\Bigl(6 \, z^{5} y^{18} + 2 \, z^{4} y^{18} - 10 \, z^{5} y^{16} + 6 \, z^{4} y^{16} + 3 \, z^{5} y^{14} - 11 \, z^{4} y^{14} - 2 \, z^{3} y^{14}\\
& + 11 \, z^{4} y^{12} + 9 \, z^{3} y^{12} - 6 \, z^{4} y^{10} - 2 \, z^{3} y^{10} + 9 \, z^{2} y^{10} + 8 \, z^{3} y^{8} - 8 \, z^{2} y^{8} + 3 \, z^{3} y^{6}\\
& + z y^{8} - 7 \, z^{2} y^{6} - 3 \, z y^{6} - 9 \, z^{2} y^{4} + 7 \, z y^{4} - y^{4} + 9 \, z y^{2} - 2 \, y^{2} - 3\Bigr) z y^{4}.
\end{eqnarray*}
The denominator of $G_3(y,z)$ is
\begin{eqnarray*}
&\Bigl(3 \, z^{6} y^{20} + z^{5} y^{20} - 4 \, z^{6} y^{18} + 2 \, z^{5} y^{18} + z^{6} y^{16} - 3 \, z^{5} y^{16} - 3 \, z^{4} y^{16} + 4 \, z^{5} y^{14}\\
& + 4 \, z^{4} y^{14} - 2 \, z^{5} y^{12} - 3 \, z^{4} y^{12} + 4 \, z^{3} y^{12} + 4 \, z^{4} y^{10} - 2 \, z^{3} y^{10} + z^{4} y^{8} - 5 \, z^{3} y^{8}\\
& - 4 \, z^{3} y^{6} + 6 \, z^{2} y^{6} + 6 \, z^{2} y^{4} - 3 \, z y^{4} - 4 \, z y^{2} + 1\Bigr) {\left(z - 1\right)}^{2}.
\end{eqnarray*}
The numerator of $g_4(z)$ is
\begin{eqnarray*}
-&\Bigl(84 \, z^{10} + 202 \, z^{9} - 72 \, z^{8} - 473 \, z^{7} + 230 \, z^{6} + 693 \, z^{5} - 393 \, z^{4} - 300 \, z^{3}\\
& + 149 \, z^{2} - 42 \, z + 10\Bigr) z.
\end{eqnarray*}
The denominator of $g_4(z)$ is
\begin{eqnarray*}
&\Bigl(28 \, z^{11} + 50 \, z^{10} - 48 \, z^{9} - 112 \, z^{8} + 140 \, z^{7} + 151 \, z^{6} - 209 \, z^{5} - 17 \, z^{4}\\
& + 66 \, z^{3} - 45 \, z^{2} + 13 \, z - 1\Bigr) {\left(z - 1\right)}^{2}.
\end{eqnarray*}
The numerator of $g_5(z)$ is
\begin{eqnarray*}
&-\Bigl(4 \, z^{30} - 48 \, z^{29} - 278 \, z^{28} + 2950 \, z^{27} + 13852 \, z^{26} - 43774 \, z^{25} - 401310 \, z^{24} - 252986 \, z^{23}\\
& + 3419602 \, z^{22} + 3192142 \, z^{21} - 10681356 \, z^{20} - 11646759 \, z^{19} + 15963209 \, z^{18}\\
& + 20876587 \, z^{17} - 13792528 \, z^{16} - 27136640 \, z^{15} + 11141815 \, z^{14} + 28709449 \, z^{13}\\
& - 17089827 \, z^{12} - 15148184 \, z^{11} + 16209330 \, z^{10} - 128310 \, z^{9} - 5691052 \, z^{8} + 2506529 \, z^{7}\\
& + 117553 \, z^{6} - 439661 \, z^{5} + 166496 \, z^{4} - 32204 \, z^{3} + 3889 \, z^{2} - 325 \, z + 15\Bigr) z.
\end{eqnarray*}
The denominator of $g_5(z)$ is
\begin{eqnarray*}
&\Bigl(2 \, z^{31} - 30 \, z^{30} - 34 \, z^{29} + 1368 \, z^{28} + 2688 \, z^{27} - 23089 \, z^{26} - 119962 \, z^{25} + 93428 \, z^{24}\\
& + 1051157 \, z^{23} + 82962 \, z^{22} - 3336331 \, z^{21} - 1267655 \, z^{20} + 4915006 \, z^{19} + 3164355 \, z^{18}\\
& - 4144167 \, z^{17} - 5690282 \, z^{16} + 3952819 \, z^{15} + 6904233 \, z^{14} - 6147688 \, z^{13} - 3023145 \, z^{12}\\
& + 5346947 \, z^{11} - 1129508 \, z^{10} - 1544617 \, z^{9} + 1124479 \, z^{8} - 183068 \, z^{7} - 122213 \, z^{6}\\
& + 84913 \, z^{5} - 26274 \, z^{4} + 4945 \, z^{3} - 580 \, z^{2} + 38 \, z - 1\Bigr) {\left(z - 1\right)}^{2}.
\end{eqnarray*}

\bibliographystyle{iopart-num}
\bibliography{biblio}

\end{document}